# A new dual RF sensor in gas detection and humidity influence

Julien George[1], Hamida Hallil[2,3], Corinne Dejous[2], Eric Cloutet[4], Aurelien Périgaud[1], Stéphane Bila[1], Dominique Baillargeat[1]

(1): CNRS, XLIM UMR 7252, University of Limoges, Limoges, France
(2): Bordeaux INP, CNRS, IMS, UMR 5218, Univ. Bordeaux, Talence, France
(3): UMI 3288 CINTRA CNRS-NTU-THALES, Nanyang Technological University, Singapore
(4): LCPO, UMR 5629, ENSCBP, IPB, Université de Bordeaux, Pessac, France

*Abstract*—In this paper, we propose to study the influence of moisture on passive RF sensors associated with poly (3,4-ethylenedioxythiophene) polystyrene sulfonate - multi walled carbon nanotubes (PEDOT: PSS - MWCNT) as a sensitive material. The designed sensor consists of a copper ring resonator and two PEDOT:PSS-MWCNT stubs located in the maximum electric fields of the fundamental resonant mode operating at 3.94 GHz. For performing a differential analysis, experiments are conducted by considering the resonant structure with and without a sensitive layer. Our experiments show the high interference of humidity on the sensor's performances and highlight the importance of differential sensing to decorrelate physical and chemical surrounding parameters. This differential innovative structure paves the way for a very high sensitive passive radio frequency (RF) sensor dedicated for gas sensing.

## I. Introduction

The monitoring of the air pollution and environmental issues have become an international priority in recent years due to its economic, ecological and health impact, with nearly 6.5 million premature deaths each year [1]. Being able to measure accurately toxic gas and Volatile Organic Compounds (VOCs) is an essential aim in this fight against pollution.

Furthermore, the development of the Internet of Things (IoT) is one the promised solution since the control of those analytes, in addition to temperature, pressure and humidity, is possible with wireless solutions and well-developed networks. However, this envisioned paradigm requires low-cost and energy-efficient technologies. Radio Frequency (RF) resonators are most important devices for the development of this market. Especially, microwave transducers associated with selective and sensitive materials will be the key of reaching the goal of performant gas sensor suitable to the future envisioned smart cities.

The use of carbon, oxides and polymers materials are important in passive RF sensors [2]–[4]. They are selected according to the target analyte, which consequently modifies their electrical properties. This change impacts the overall electrical behavior of the RF structure, resulting in a shift of the resonant frequency of the resonator and then allows the quantification of very low concentration of the target species.

PEDOT:PSS is a polymer consisting of poly(3,4-ethylenedioxythiophene) (PEDOT) and sodium poly(styrene sulfonate) (PSS). The former is a conjugated polymer derived from polythiophene in which a fraction of the sulfur atoms is protonated with a positive charge. The second is a negatively charged sulfonated polystyrene in which a fraction of the sulfonate groups $SO_3^-$ carries a sodium ion $Na^+$.

PEDOT:PSS-MWCNT is a common sensitive material associated with transducers to perform sensors dedicated for gas and VOCs detection [5]–[7]. However, the consideration of humidity in interaction with this material usually is overlooked in a large amount of academic work knowing that is the first interferent to be present in real atmospheric detection.

In this paper, simulations and measurement of the microwave sensor are illustrated and reported. Then the exposition of the sensor under different concentration of humidity and dry atmosphere are discussed and analyzed.

## II. Sensor Design and Validation

### A. Resonator design

The proposed sensor geometry is a ring resonator operating in reflection. It is a half-wave structure having, on its fundamental mode, two locations of maximum electric fields where sensitive stubs composed of PEDOT:PSS-MWCNT are placed, as depicted in Fig. 1. Its original design stems from our need to reduce the size though ensuring a low resonance frequency. The areas where the electric field is weak are folded in order to minimize the effects of this miniaturization.

The resonator, made of copper, thus forms a 14x14 mm² square with a track width of 2.5 mm. The PEDOT:PSS-MWCNT synthetized by PolyInk is deposited by inkjet and forms rectangles of 4.5 mm long by 2.5 mm wide placed parallel to the ring. The substrate used is a 100 μm thick polyimide with a relative permittivity of 3.5 with loss tangents of 0.0127 at 2.4 GHz. PEDOT:PSS has been characterized at $66.10^{-3}$ S.m$^{-1}$ in previous work [5].

For our experiment, two devices are used at the same time. The resonator with the sensitive material is referred to as the "sensitive channel" while the second bare resonator is referred to as the "reference channel". The goal of the reference is to compensate for physical parameters influence such as temperature and pressure on the sensor, that can impact similarly both the channels. Thus, by performing differential analysis it is possible to study only the effect of the target analyte which is not usually considered in previous studies [8].

While Fig. 1a illustrates the sensitive and reference channels, Fig. 1b represents the distribution of the electric field across the sensitive structure. The warm colors highlight the concentration of the electric field in the PEDOT: PSS-MWCNT stubs.

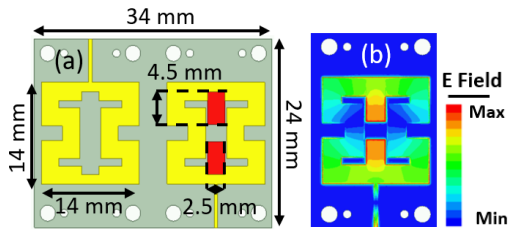

Fig. 1: (a) Schematic of the reference (left) and sensitive (right) channels; (b) Simulated electrical field of the sensitive channel

Fig.2 shows the simulated Scattered (S) parameter response ($S_{11}$: reflection), in magnitude and phase, of the sensitive resonator over a large frequency band from 3.5 GHz to 4.5 GHz. In addition, this figure illustrates the behavior of the sensor with three PEDOT:PSS-MWCNTs conductivities, close to the value characterized in [5], between 45 and 55 kS.m$^{-1}$. Indeed, the interaction of $H_2O$ molecules with the sensitive material should increase the conductivity of the PEDOT: PSS polymer material. From this simulation, it can be observed a diminution of the resonant frequency of the sensitive channel in the presence of moisture. The related sensitivity is estimated to – 2.14 MHz.(%cond)$^{-1}$.

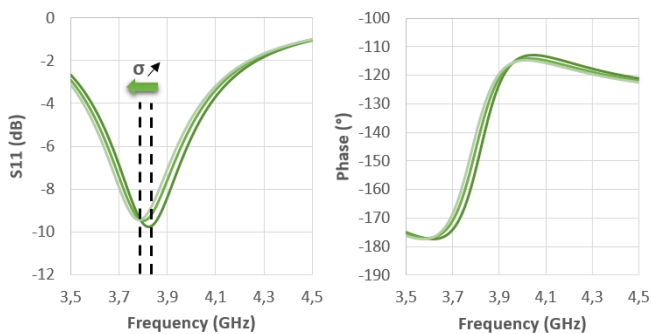

Fig. 2: Parameter $S_{11}$ of the sensitive channel in function of PEDOT: PSS-MWCNT conductivity

### B. Fabrication and validation of the sensor

The resonant rings have been manufactured by BetaLayout, using copper and polyimide materials with a process derived from their typical flexible Printed Circuit Board. Briefly, a copper layer as well as a photosensitive resin are laminated on the flexible substrate. The pattern is obtained by localized laser exposure, followed by a development to remove the insolated resin and an etching of the bare copper.

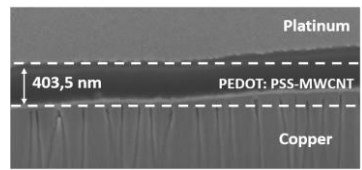

Fig. 3: Thickness measurement of PEDOT:PSS-MWCNT by SEM (with top platinum to avoid charging effects)

The PEDOT: PSS-MWCNT supplied by Poly-ink is ink jet deposited by the Center for Technology Transfers in Ceramics (CTTC). The thickness measurement of PEDOT:PSS-MWCNT stubs was carried out using a Scanning Electron Microscope (SEM) as shown in Fig. 3. It was estimated to be equal to 0.4 µm.

Measurements under ambient conditions give for the sensitive channel a resonance frequency of 3.94 GHz and 4.03 GHz for the reference channel (Fig. 4), close to the simulated values (Fig. 2). The presence of a conductive stub on the sensitive channel increases the effective length of the structure, which in turn decreases the resonance frequency, compared to the reference channel. It can be noted also the greater quality factor of the sensitive channel due to an optimization focused on this path.

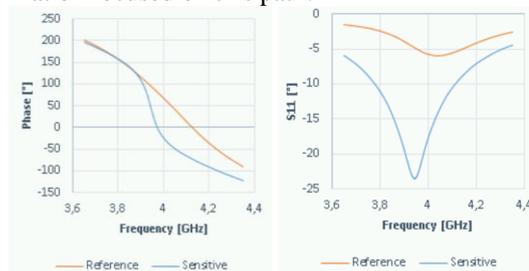

Fig. 4: Electrical characterization ($S_{11}$ parameter) of sensitive (blue) and reference (orange) channels

### III. EXPERIMENTAL SETUP

In some cases, prior to humidity tests, a treatment of the sensors in oven at 80°C for 10 min was performed to minimize the humidity adsorbed in the materials, especially at the surface of the PEDOT: PSS-MWCNT.

Then gas test measurements are carried out using a hermetic cell with a flow inlet and outlet. The humidity is transported by nitrogen, the relative humidity (%RH) level and the total flow are controlled by the vapor generator CALIBRATION PUL 110. The electrical measurement is performed using an Anritsu MS2026B vector network analyzer (VNA) from 3.65 to 4.35 GHz with a resolution of 175 kHz (Fig. 5). A computer controls acquisition every 24 seconds.. Finally, the data are processed. All measurements have been done at room temperature.

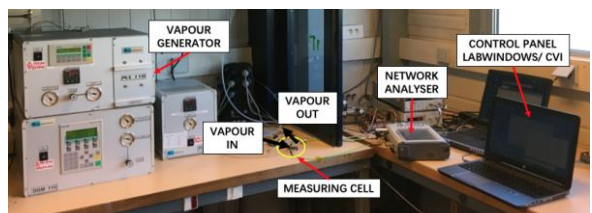

Fig. 5: Experimental setup for sensor characterization with humidity

The measurement sequences performed in this study consist first in generating a dry nitrogen step of 180 minutes to decrease humidity to 4.4 %RH in the cell and remove impurities on the surface of the device. Second, a succession of nitrogen and humidity sequences of 10 minutes each are generated for a complete experiment duration of 100 minutes. The humidity concentration sequence are 28 %RH (two times), 35 %RH (two times) and finally 45 %RH is presented in Fig. 7.

## IV. RESULTS AND DISCUSSION

The results with a first sample without any heat pre-treatment are presented in Fig. 6 and Fig. 7. In Fig. 6, the overlay of all $S_{11}$ parameter measurements shows more dispersion with the sensitive channel than with the reference one, which is expected from the additional response inherent to the sensitive material response to humidity. In Fig. 7, we illustrate the post-processed real-time resonance frequency of the sensitive (dashed black line) and reference (grey line) channels. From these results, we observe a repeated decrease of the resonance frequency of 6 MHz with the sensitive channel when the humidity measured in the cell changes from 4.4 %RH to 28 %RH. From that, the sensitivity of our structure could be estimated of 214 kHz.%RH$^{-1}$ and the response time is 1 minute and 12 seconds. This is in agreement with an expected increase of the PEDOT:PSS-MWCNT conductivity with the concentration of humidity, as simulated in Fig. 2, which would lead to an increase of the conductivity of 3%. On the contrary, no significant shift is visible on the reference channel, except a continuous drift of –1 MHz after 100 minutes. This is consistent with the fact that the sensor response to humidity is mainly attributed to the sensitive coating.

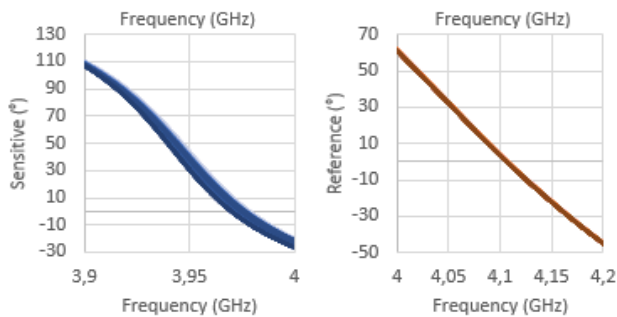

Fig. 6: Superposition of the $S_{11}$ parameter of the sensitive channel (blue curves, left) and reference one (red curves, right), over the 100-minutes humidity vapor sequence

Furthermore, it must be noted also from Fig. 7, that the response of the sensitive channel is approximately the same at 28%, 35% or 45% of humidity, which could be explained by a rapid saturation of the PEDOT: PSS-MWCNT.

Indeed, PEDOT: PSS films are known to strongly depend on humid environment due to the hygroscopic nature of PSS which plays both the role of counterion and aqueous dispersing agent. In fact, depending on the ratio between PEDOT and PSS, the ambient humidity can have opposite effects on the film's conductivity [9]. We are here in the case where PSS part swells in the presence of water vapor favoring the segregation and percolation of PEDOT nanodomains thus increasing the conductivity of the PEDOT film. This is in agreement with the saturation phenomenon that is observed at high humidity % due to the excess minority hydrophilic PSS.

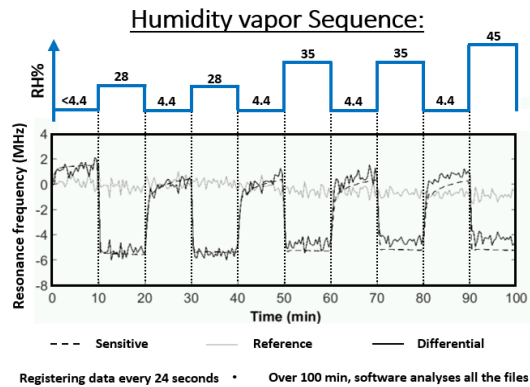

Fig. 7: Humidity vapor sequence and real-time resonance frequency measurement, sample without heat pre-treatment

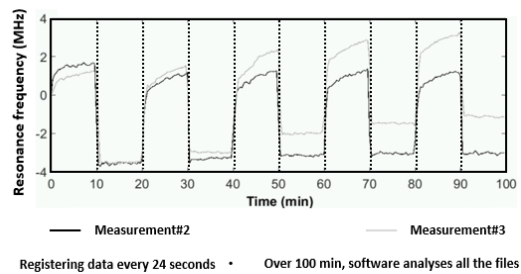

Fig. 8: Real-time resonance frequency of sensitive channel over the humidity vapor sequence, sample with heat pre-treatment

Similar experiments with heat pre-treatment led to results presented in Fig. 8. They confirm the previous results, both the sensitivity with frequency shifts close to – 6 MHz too, and saturation effect. They also demonstrate the repeatability of our measurements. Considering the frequency shift obtained from 4.4 %RH to 28 %RH, and already reaching the saturation domain, it may be inferred a sensitivity superior to 250 kHz.%RH$^{-1}$ at low %RH. This is a huge value that may be useful for some applications. The other interesting fact is the saturation domain itself. Indeed, in many environments, the ambient humidity is superior to 30 %RH. Thus, it could be interesting for sensing applications, that a change of humidity, while remaining superior to this value, does not influence the sensor, provided the sensitive coating still interacts with the target species for which it has been designed.

In addition, we observe an increase of the drift over time, especially on the third experiment. That may be explained by a modification and degradation of the substrate polymer materials at the atomic scale organization.

## V. CONCLUSION

This study allowed to put into evidence a new dual RF structure as chemical sensor, with PEDOT: PSS-MWCNT as sensitive material on one of both channels, the bare one acting as a reference to compensate for parameters such as the temperature. We focused on the sensitivity to humidity, which is the main potential interferent in the environment. The results highlighted a very high sensitivity to humidity, with a response time of 1 minute and 12 seconds.

Measurements are repeatable on the same structure and verified on several devices. Finally, the PEDOT:PSS-MWCNT is saturated at humidity levels of 28 %RH or less, which means that it should not be affected by the humidity variations that can occur in most of real conditions. However, our study shows that nitrogen flow can reduce the humidity level sufficiently to influence experimental results in the laboratory.